\documentclass[11pt]{article}
\usepackage{amssymb}

\def\re#1{(\ref{#1})}

\def\beeq{\begin{eqnarray}}  
\def\beeqn{\begin{eqnarray*}}  
\def\eeeq{\end{eqnarray}}  
\def\eeeqn{\end{eqnarray*}}  

\def\bit{\begin{itemize}}  
\def\eit{\end{itemize}}  
\def\ben{\begin{enumerate}}  
\def\een{\end{enumerate}}  
  
\newcommand{\beq}{\begin{equation}} \newcommand{\eeq}{\end{equation}}  
\newcommand{\bea}{\begin{eqnarray}} \newcommand{\eea}{\end{eqnarray}}  
\newcommand{\ber}{\begin{array}} \newcommand{\eer}{\end{array}}  
  
\newcommand{\dms}{\displaystyle}  
\newcommand{\dx}{d^{4}x}                \newcommand{\de}{\partial}  
\newcommand{\dy}{d^{4}y}                  
            \newcommand{\fp}{\Phi\Pi}   
        
\newcommand{\alp}{\alpha}             
\newcommand{\nabh}{\hat{\nabla}}

                \newcommand{\Om}{\Omega}  
          \newcommand{\gam}{\gamma}    
                 
\newcommand{\bet}{\beta}                \newcommand{\del}{\delta}

 \newcommand{\g}{{\bf  \Gamma}}

\def\h#1{\hat{#1}}   
\def\b#1{\overline{#1}}

\def\sla#1{\not \!\!{#1}}  
\def\oop#1{\vspace{#1}}

\def\dfa#1#2#3{\frac{\delta}{\delta {#1}^{#2}_{#3} }}  
\def\dfu#1#2{\frac{\delta}{\delta {#1}^{#2} }}

\def\fd#1#2#3#4{\dms{\frac{\delta #1}{\delta {#2}^{#3}_{#4}}}}  
\def\fdu#1#2#3{\dms{\frac{\delta #1}{\delta {#2}^{#3}}}}  
\def\fdd#1#2#3{\dms{\frac{\delta #1}{\delta {#2}_{#3}}}}

\begin{document}  
\begin{flushright} 
MPI/PhT/97-89  \\
hep-th/9804013 \\
December 1997
\end{flushright}  
 
\def\bc{\bar{c}}  
\def\gg{\widetilde{\g}}  
\def\gc{\g_{0}}   
 
\vspace{2cm} 
\begin{center} 
{\Huge \rm  The Abelian Anti-ghost Equation 
\\ for the Standard Model in the \vspace{.3cm}  \\
't Hooft-Background Gauge} 

\vspace{1cm}

{\large \bf {Pietro Antonio Grassi\footnote{e-mail:pgrassi@mppmu.mpg.de}}}

{\it Max-Planck-Institut f\"ur Physik, F\"ohringer Ring 6, 80805, M\"unchen, Germany}
\end{center} 

\vspace{2cm} 
\begin{center}
{\bf  Abstract} 
\end{center}

{In this paper we study the Abelian Anti-ghost equation for the 
Standard Model quantized in the 't Hooft-Background gauge. We 
show that this equation assures the non-renormalization of the 
abelian ghost fields and prevents possible abelian anomalies.}

\vfill
\eject

\section{Introduction}

As is well known for a chiral gauge model with spontaneous symmetry breaking 
(in particular for the Standard Model (SM))   
the algebraic renormalization program (\cite{bec}, \cite{algeb}, \cite{pig}) 
seems to be the only feasible technique at our 
disposal to define the 
counterterms and to restore order by order the symmetries. In fact the 
absence of a symmetric renormalization scheme 
forces us to introduce proper counterterms which cancel any  
breaking terms (if there is no anomaly ) 
of those functional identities which express the symmetry content of the model at the 
quantum level.  

For the SM with three generations of fermions (and in presence of direct 
CP violation induced by means of the CKM matrix \cite{CP}) the 
purely algebraic treatment of counterterms is not sufficient essentially for two 
reasons: 1) new local counterterms could appear modifying the couplings 
among fermions and abelian gauge fields and 2) new anomalies could 
show up. Both problems were considered in the pioneering 
work by Bandelloni, Becchi, Blasi and Collina \cite{henri},\cite{band}  and recently 
formulated in a more general context by Barnich {\it et al} in \cite{henn}. 

The former problem expresses the fact that the BRST symmetry \cite{bec} 
for a non-semi-simple 
gauge group  does not identify uniquely the fermion and scalar representations 
for abelian factors of the group. As a consequence there arises the possibility 
that renormalized representations might be inequivalent to those of the tree approximation 
(instability of fermion and scalar representations of abelian factors). In this 
situation the definition of the charge of particles and the 
definition of photon gauge field have to be modified order by order according 
to the renormalized representations. 

On the other side new anomalies for the BRST symmetry could appear, in particular 
anomalies which depend on the BRST sources (see for example \cite{henn} for a complete 
exposition) and anomalies of the form: abelian ghost fields $\times$ 
BRST invariant polynomials (up to total derivatives).    

Both these problems can be avoided by using the Ward-Takahashi Identities (WTI) 
for QED part and by means of the CP symmetry. 
But both WTI and CP invariance cannot be used for the SM. In fact to 
avoid IR problems for massless scalar fields the `t Hooft gauge 
fixing is chosen and this leads to certain unavoidable couplings among 
ghost fields and scalar fields. In this situation the QED-WTI has to 
be replaced by the Slavnov-Taylor Identities (STI) and one runs into the above 
problems. Furthermore without the CP symmetry 
it is not trivial to prove the absence of the second type anomalies. 

There are two ways to solve these problems and to define 
correctly the SM at the higher orders: 1) introducing a new QED-WTI 
which can be pursued to higher orders ( see for example the paper by 
E. Kraus \cite{krau_ew}) or 2) defining an abelian anti-ghost 
equation (see for examples the paper \cite{henri} and \cite{band}) 
which can be established to higher orders and which provide 
non-renormalization properties for abelian ghost fields and 
for the fermion and scalar representations of abelian factors. 

According to the first alternative 
the new QED-WTI is implementable only by introducing 
new scalar classical fields coupled to ghost fields. On the other hand in the 
Background Field Method (BFM) (\cite{bkg} and \cite{msbkg}) for the SM those classical scalar fields are 
naturally introduced in order to establish the WTI for the background gauge invariance 
(both for the $U(1)$ and $SU(2)$ factors). 

Moreover by means of the first alternative only the instability of 
abelian factors and anomalies depending on the BRST sources can be avoided, and 
the second type anomalies could be still present. In contrast to this in the BFM an Abelian 
Anti-ghost Equation (AAE) 
can be defined and this excludes both instability of representations 
and any kind of anomalies. Furthermore this equation is deeply related to 
the QED-WTI which can be established by computing the (anti)commutator between 
the AAE and the  STI. 

Furthermore we want to underline that the control of instability of the 
representations of abelian factors and the vanishing of new anomalies have to be 
taken into account for the renormalizability in the ``modern'' sense as 
proposed in \cite{ren_mode}. The BFM with the AAE is a promising 
approach to manage these problems. 
  
In the present paper we will define the AAE within the BFM for a general 
non-semi-simple gauge model. 
We will discuss the non-renormalization properties 
of the representations for abelian factors 
(section 2 and section 3) and we will analyze these problems 
in the SM with three generations of fermions (section 4).   

In particular we will  
give a short account on the renormalization of Green's function with external 
BRST sources terms (section 5) since up to now there is no complete analysis 
of this point and because of their relevance in higher loop computations. The 
complete discussion will be presented in future papers. 

The appendix A deals with a summary of conventions, explicit forms of STI and of  
Faddeev-Popov equations. In the appendix B a discussion of the renormalization of 
Faddeev-Popov equations and AAE is given taking into account the IR problems.

\section{Instabilities and Anomalies} 

At the quantum level the Slavnov-Taylor Identities (STI) for the one particle irreducible 
generating functional $\g$ 
\beq\label{st_0}
{\cal S}(\g) = 0 
\eeq
are implementing the BRST transformations \cite{bec} of the classical gauge model 
guaranteeing gauge invariance and the independence of physical 
quantities of the gauge fixing parameters. Furthermore the STI single out the physical 
subspace of the space of states on which a unitary S-matrix can be defined \cite{kugo}. 
With respect to these identities the tree level effective action 
$\g_0$ is then defined as the most general invariant local polynomial compatible with the 
power counting. 

For non-semi-simple gauge models the 
general solution  with zero Faddeev-Popov charge 
of the constraints\footnote{We have to recall that 
besides to the STI, some supplementary functional equations have to be 
taken into account; in the text we will discuss one of these 
identities and in the appendix the remaining ones.} 
is found in the papers by Bandelloni {\it et al.} \cite{henri}, \cite{band} 
and by Barnich {et al.} \cite{henn}
\bea\label{ge_so}
\g_0 & = & \g^{Stab}_0 + \g^{Inst}_0 \nonumber  \\
\g^{Stab}_0 & = & 
\dms{\int \dx}\, {\cal L}(x)  +   {\cal S}_{\g_0} \dms{\int \dx} \, {\cal K}(x) \\
\g^{Inst}_0 & = & \sum_{\alp} k^{\alp}_{a_A}  \dms{\int \dx}  
\left( j^{\mu}_{\alp} A^{a_A}_{\mu} + P^{a_S}_{\alp, \mu} \gamma^{\mu}_{a_S} c^{a_A} + 
 P^{i}_{\alp}  \gamma_i c^{a_A} + \bar{P}^{I}_{\alp}  \eta_I c^{a_A} + 
P^{I}_{\alp} \bar{\eta}_I c^{a_A} 
\right)  \label{insta}
\eea
where ${\cal L}(x)$ is a general gauge invariant polynomial with Faddeev-Popov charge zero and 
dimension $\leq 4$,  
${\cal K}(x)$ is a general polynomial with Faddeev-Popov charge -1 and 
${\cal S}_{\g_0}$ is the linearized Slavnov-Taylor operator.   
$k^{\alp}_{a_A}$ are arbitrary constants, $A^{a_A}_{\mu}$ are the abelian 
gauge fields and $ c^{a_A}$ their corresponding ghost fields; 
as described in the appendix $\gamma^{\mu}_{a_S}, \gamma_i, \eta_I , \bar{\eta}_I$ 
are the external sources for the BRST variations of the gauge fields 
 $A^{\mu}_{a_S}$, of the scalars $\phi_i$, and of the fermions $\psi_I, \bar{\psi}_I$. 

Finally we suppose the existence of some rigid symmetries for the 
invariant action $\g^{Inv}_0 = \int \dx {\cal L}(x)$ and their 
corresponding BRST-invariant currents  
\beq\label{co_cu}
\de_{\mu}  j^{\mu}_{\alp} =  P^{a_S}_{\alp, \mu} \fd{\g^{Inv}_0}{A}{a_S}{\mu} + 
P^{i}_{\alp} \fdu{\g^{Inv}_0}{\phi}{i} + \bar{P}^{I}_{\alp} \fdu{\g^{Inv}_0}{\bar{\psi}}{I} + 
\fdu{\g^{Inv}_0}{\psi}{I} P^{I}_{\alp}
\eeq
where $P^{a_S}_{\alp, \mu}, P^{i}_{\alp},  \bar{P}^{I}_{\alp},  P^{I}_{\alp}$ are 
local polynomials for each $\alp$. 
  
Since the proof of the existence and uniqueness of 
terms  $\g^{Inst}_0$ is given in \cite{henn}, we can simply observe that  
by acting with the Slavnov-Taylor operator ${\cal S}_{\g_0}$ on the 
$\g^{Inst}_0$ we get 
\begin{eqnarray}\label{so_in}
\hspace{-1cm} {\cal S}_{\g_0} \left( \g^{Inst}_0 \right) & = & 
\sum_{\alp} k^{\alp}_{a_A} \dms{\int} \dx \left\{ 
\left[ 
\left( {\cal S}_{\g_0} j_{\alp}^{\mu} \right) A^{a_A}_{\mu} + 
j^{\alp}_{\mu} \de^{\mu} c^{a_A} \right] +   
c^{a_A} \left[ P^{a_S}_{\alp, \mu} {\cal S}_{\g_0} \gamma^{\mu}_{a_S} +  
\left( {\cal S}_{\g_0} P^{a_S}_{\alp, \mu} \right) \gamma^{\mu}_{a_S} + 
\right. \right. \nonumber  
\\ && \left. \left. + 
P^{i}_{\alp} {\cal S}_{\g_0}  \gamma_i +  
\left( {\cal S}_{0} P^{i}_{\alp} \right) \gamma_i + 
\bar{P}^{I}_{\alp} {\cal S}_{\g_0} {\eta}^{I} + \left( 
{\cal S}_{\g_0} \bar{P}^{I}_{\alp} \right) {\eta}^{I}  + 
\b{\eta}_I {\cal S}_{\g_0} P^{I}_{\alp}  +   
\left( {\cal S}_{\g_0}\bar{\eta}_I \right) 
P^{I}_{\alp} \right] \right\} 
\end{eqnarray}

Making explicit the action of the Slavnov-Taylor operator on the 
BRST sources, 
${\cal S}_{\g_0}\bar{\eta}_I = \fdd{\g_0}{\psi}{I}, {\cal S}_{\g_0}  
\gamma^{\mu}_{a_S} = 
\fd{\g_0}{A}{a_A}{\mu},  {\cal S}_{\g_0}  \gamma_i = \fdd{\g_0}{\phi}{i}$, using the 
invariance of the conserved currents $j_{\alp}^{\mu}$, 
and integrating by parts we obtain
\begin{eqnarray}\label{so_in_2} 
{\cal S}_{\g_0} \left( \g^{Inst}_0 \right) = 
\sum_{\alp} k^{\alp}_{a_A} \dms{\int \dx }  c^{a_A} \left[ - \de^{\mu}
j^{\alp}_{\mu} +  P^{a_S}_{\alp, \mu} \fd{\g^{Inv}_0}{A}{a_A}{\mu} + 
P^{i}_{\alp} \fdd{\g^{Inv}_0}{\phi}{i} + \bar{P}^{I}_{\alp} \fdd{\g^{Inv}_0}{\bar{\psi}}{I} + 
 \fdd{\g^{Inv}_0}{\psi}{I}  P^{I}_{\alp} 
\right]  
\end{eqnarray}

From the hypotheses of current conservation (\ref{co_cu}) we immediately get 
the invariance of $\g^{Inst}_0$.

This proves that the new terms can be added to 
the tree level effective action without violating the STI.  
The dependence on only the abelian gauge fields $A^{a_A}_{\mu}$  
expresses that the 
abelian symmetries are not protected against radiative 
corrections contrary to the non-Abelian ones. Generally 
the generators of the Abelian factor ${\cal G}_A$ of the gauge group 
${\cal G}$ can be mixed modifying the 
quantum numbers of the physical states. This instability of the representations of the 
abelian factors was  
already analyzed in the pioneering works by Bandelloni {\it et al.}  
\cite{henri} where they observe that  
only the non-renormalization properties for the 
abelian ghost couplings could prevent this phenomena.

As will be discussed in the forthcoming sections 
the presence of  $\g^{Inst}_0$ in the SM 
with three generation of fermions is a consequence of the 
 conserved quantum numbers hyper-charge, baryon and lepton numbers \cite{acci} and their 
gauge invariant currents. 

Moreover due to its relevance in the 
renormalization procedure we would like to describe briefly 
the anomalies of the BRST symmetry for 
a non-semi-simple gauge model. 
As proved in the papers \cite{henri} and \cite{henn} the structure of the general 
solution of the consistency conditions \cite{wess} is equal to the 
cohomology 
\bea
&& \hspace{-1cm} {\cal A}  = \dms{\int \dx} {\cal A}^{ABJ}(x) +   \dms{\int \dx}  
{\cal A}_{1}(x) + 
\dms{\int \dx}  {\cal A}_{2}(x) +
{\cal S}_{0} \dms{\int \dx} {\cal B}(x) \label{ano_0} \\
 &&\hspace{-1cm} {\cal A}^{ABJ}(x)  =  \sum_{i} r_i \epsilon^{\mu\nu\rho\sigma} 
\left[ D_{i}^{abc} c^{a} \de_{\mu} A^{b}_{\nu}  \de_{\rho} A^{c}_{\sigma} + 
\dms{\frac{F^{abcd}_i}{12}} ( \de_{\mu} c^{a} ) A^{b}_{\nu} A^{c}_{\rho} A^{d}_{\sigma} 
\right]  \\
 &&\hspace{-1cm} {\cal A}_{1}(x) =  c^{a_A} {\cal R}_{a_A}(x) \label{ano_1} \\
 &&\hspace{-1cm} {\cal A}_{2}(x)  =  w^{\alp}_{a_A,b_A} \left( j_{\alp}^{\mu} A^{a_A}  c^{b_A} + 
\frac{1}{2}  P^{a_S}_{\alp, \mu}  \gamma^{\mu}_{a_S} c^{a_A} c^{b_A} + 
\frac{1}{2}  P^{i}_{\alp} \gamma_i c^{a_A} c^{b_A} + 
\frac{1}{2} \bar{P}^{I}_{\alp}  {\eta}^{I} c^{a_A} c^{b_A} + h.c. 
\right) \label{ano_2}
\eea
where ${\cal A}^{ABJ}(x)$ is the well known Adler-Bardeen-Jackiw anomaly \cite{abj} and  
the $r_i$ are its coefficients; the tensors $F^{abcd}_i$ are defined by 
$$ 
F^{abcd}_i = D_{i}^{abx} (ef)^{xcd} + D_{i}^{adx} (ef)^{xbc} + D_{i}^{acx} (ef)^{xdb}
$$
with $D_{i}^{abc}$ invariant symmetric tensors of rank three 
on the algebra ${\cal G}$. ${\cal B}(x)$ in eq. (\ref{ano_0}) is a generic  
polynomial with dimension $\leq4$ and Faddeev-Popov charge zero\footnote{It provides the 
non-invariant counterterms to cancel the spurious anomalies coming 
form a non-symmetric renormalization scheme.}; ${\cal R}_{a_A}(x)$ are a set of 
BRST invariant polynomials. The latter are responsible 
for CP-violating anomalies, which might show up in the SM 
as terms like the following 
\bea\label{cp_sm}
&& \dms{\int \dx } \left(- \sin \theta_W c_Z +  \cos \theta_W c_{\gamma} \right)
\left(  (H + v)^2 + G^2_0 + 2\, G^+ G^- \right), \nonumber \\
&& \dms{\int \dx } \left(- \sin \theta_W c_Z +  \cos \theta_W c_{\gamma} \right)
\left(  (H + v)^2 + G^2_0 + 2\, G^+ G^- \right)^2,\hspace{.2cm} {\rm or} \\ 
&& \dms{\int \dx } \left(- \sin \theta_W c_Z +  \cos \theta_W c_{\gamma} \right) 
{\cal F}_{\mu\nu}^{a} {\cal F}^{\mu\nu}_{a} \nonumber 
\eea
where $H,G_0, G^{\pm}$ are respectively the Higgs field, the neutral would-be Goldstone 
field and the charged would-be Goldstone fields; $c_Z,c_{\gamma} $ are the ghost field for the 
Z gauge boson and for the photon, ${\cal F}_{\mu\nu}^{a}$ is the field strength tensor 
of the gauge fields $A^{a}_{\mu}$ 
and, finally, $\theta_W$ is Weak mixing angle. 
They are absent if the discrete CP symmetry is valid.  

In the expression ${\cal A}_2$ of eq. (\ref{ano_2})
the coefficients $w^{\alp}_{a_A,b_A}$ are 
constant and antisymmetric in the abelian indices $a_A,b_A$ for each $\alp$.  

As observed by Barnich {et al.} \cite{henn} the anomaly terms 
(\ref{ano_2}) are trivial if and only if the conserved currents  $j_{\alp}^{\mu}$ 
are trivial, that is, are equal on-shell to an identically conserved total divergence; however 
as already observed, in the SM, the conserved lepton and baryon numbers  
provides four non-trivial examples of $j_{\alp}^{\mu}$. About the actual  
presence of anomalies as (\ref{ano_1}) and (\ref{ano_2}) at higher orders  there is no 
evidence from one and two loop calculations 
and  the only example where (\ref{ano_1}) occurs in the literature is given in 
\cite{medrano}. 
In fact by choosing an appropriate gauge fixing, one can avoid these anomalies 
as in the `t Hooft gauge as proved in \cite{henri} or 
as in the `t Hooft-Background gauge as proved in the present 
paper. 

In 't Hooft gauge the abelian ghosts 
are coupled to scalars in order to protect the model against IR divergences
of massless would-be Goldstone fields. As a consequence  anomalies as
${\cal A}_1, {\cal A}_2$ might appear.  
However the abelian ghosts    
couple to the quantized fields only by means of super-renormalizable 
vertices (with the exception of the BRST source terms), hence 
they decouple at high momenta. By Weinberg's theorem \cite{wein},  
the coefficients of these anomalies are vanishing. 
On the other hand for non-linear gauge fixing as {\it e.g.} Fujikawa's type ( see
for example \cite{fuji} and \cite{appl_fuji}) there are  
non trivial hard couplings for the abelian ghosts  and no non-renormalization
property assures their decoupling. 

In order to extend the asymptotic decoupling of abelian ghosts in
't Hooft gauge to all orders,  
Bandelloni {\it et al.} in \cite{henri} add {\it ad hoc}    
a new  term coupled with a field with dimension two 
to the classical Lagrangian leading to an anti-ghost equation which provides the 
wanted non-renormalization properties.   
Here, we will show that 
in the BFM this equation arises automatically because of the    
presence of the sources $\Om_i$ coupled to certain composite   
operators. This equation excludes {\cal a priori} terms like    
(\ref{insta}) and any hard coupling of the Abelian ghost fields $c^{a_A}$ up to    
BRS source terms.

\section{The Abelian Anti-ghost Equation in the `t Hooft-background gauge}   

As discussed above the choice of the    
't Hooft-like gauge fixing introduces couplings for the Abelian ghost fields.   
However in the BFM we have to our disposal the abelian anti-ghost 
equation which leads to an abelian WTI. In this section we will 
derive the abelian anti-ghost equation and, finally, we will show 
how the abelian WTI is related to the   
abelian anti-ghost equation.

The abelian WTI it is sufficient to exclude the instability terms (\ref{insta})
guaranteeing the absence of mixing among the abelian generators. However, it is
not sufficient to assure the absence of anomalies as (\ref{ano_1}) and only
the AAE can solve\footnote{In the polynomials ${\cal R}_{a_A}$ both hard and
soft anomalies are contained. In paper \cite{henri} and \cite{bec} an argument
is presented to prove that if the hard anomalies, i.e. the higher dimension
terms in  ${\cal R}_{a_A}$, are absent then, by the Callan-Symanzik
equation, also the lower dimension terms ( as the first example
in the (\ref{cp_sm})) vanish.} this problem.

Firstly we want to explain how to derive an anti-ghost 
equation in a general non-semi-simple gauge model. As
is well known, the Faddeev-Popov terms
for a gauge model are given by constructing a BRST invariant gauge  fixing function
\cite{bec}. If the gauge fixing is a linear functional in 
the gauge bosons and the scalars as in the 
't Hooft-like gauge, we get trilinear couplings for
the ghost fields with the gauge bosons and  
the scalars.  
Then the equation of motion for ghosts are non-trivial and their
validity at the quantum level requires the renormalization of 
new composite operators. Moreover the ghost fields $c^a$ and
anti-ghost fields $\bc^a$ describe independent degrees of freedom. In fact, for the equation 
of motion of the $\bc^a$
the necessary composite operators are already contained in the
tree level action coupled to BRST sources ( see below in the eq. (\ref{sou}))
and they require no independent renormalization, while 
for the $c^a$ the corresponding operators are absent  and
one is obliged to introduce them with their proper renormalization. 
This problem is solved in the BFM because the  operators needed for the
abelian anti-ghost equation are already included in the tree level action.  

The relevant parts of the action which enter in the present derivation are the 
gauge fixing terms, the Faddeev-Popov action and the BRST source terms.  
In the 't Hooft-background gauge they are (see also
the papers by Denner {\it et al.}\cite{msbkg} for the 't Hooft-background gauge fixing
in the physical field representations and 
one loop computations ):
\bea\label{gaufix}  
&& \g_0 = \g^{Inv}_0 +
\dms{\int} \dx {\cal L}^{g.f.}(x) +
\dms{\int} \dx {\cal L}^{\fp}(x) + 
\dms{\int} \dx {\cal L}^{S.T.}(x) \nonumber \\
&&\hspace{-1cm} {\cal L}^{g.f.} = 
    b^{b}  
    \left[\del^{ba_{S}} \nabh ^{a_{S}b_{S}}_{\mu} (A-\h{A})^{\mu}_{b_{S}} + 
    \del^{ba_{A}} \partial_{\mu} (A-\h{A})^{\mu}_{a_{A}} +  
    \rho^{bc} (\h{\phi}+v)_{i} (et)^{c}_{ij} (\phi+v)_{j} + 
    \frac{\dms{\Lambda^{bc}}}{2} b_{c}  \right]   \nonumber \\
&&\hspace{-1cm}
{\cal L}^{\fp} = - \bc_{a} \left[\del^{aa_{S}} 
    \nabh^{a_{S}b_{S}}_{\mu} \nabla_{\mu}^{b_{S}c_{S}} c_{c_{S}} + 
    \del^{aa_{A}} \partial^{2} c_{a_{A}}  + 
    \rho^{ab} (\h{\phi}+v)_{i} (et)^{b}_{ij} (et)^{c}_{jk}(\phi+v)_{k} c_{c}  +  
  \right. \nonumber \\ 
&&\left.
    + \nabla_{\mu}^{a_{S}b_{S}} \Om^{b_{S}}_{\mu} + 
    \rho^{ab} \Om_{i} (et)^{b}_{ij} (\phi+v)_{j}  \right] \label{gua_fi_2} \\
&&\hspace{-1cm}
{\cal L}^{S.T.} = 
    \gam_{a_{S}}^{\mu} 
    \left( \de_{\mu} c_{a_{S}} - 
      (ef)^{a_S b_S c_S} A_{\mu}^{b_{S}} c_{c_S} \right) + 
    \gam^{i} 
    \left( c_a (et)^a_{ij} (\phi_{j} + v_j)\right) +  \nonumber \\ 
&& +
      \zeta^{a_{S}} 
      \left( -\frac{1}{2} (ef)^{a_S b_S c_S} c_{b_{S}} c_{c_S} \right) + \b{\eta}_{I} 
        c^{a}(eT)^{a}_{IJ} \psi_{J} +  \b{\psi}_{I} 
        c^{a}(eT)^{a}_{IJ} \eta_{J} \label{sou}
 \eea            
where $\g^{Inv}_0$ is the invariant action of the non-semi-simple gauge model 
coupled to fermions and scalar fields
in the representations discussed in appendix A (where 
also the definitions of covariant derivatives  are given). The symmetric invariant 
constant matrices 
$\rho^{ab}, \Lambda^{ab}$ 
are respectively the 't Hooft parameters and the gauge 
fixing parameters.  By construction the gauge fixing\footnote{
For explicit computations of radiative corrections \cite{msbkg}
it is convenient to introduce the  
background gauge fields for the abelian gauge bosons ${A}^{a_A}_{\mu}$.
It can be proved that those background fields are unessential, in fact they
are related to the abelian Nakanishi-Lautrup fields $b^{a_A}$
by $$\fd{\g}{A}{a_A}{\mu} = \de_{\mu} b^{a_A}.$$ 
This equation has to be compared with the analogous equations 
for the non-abelian background gauge fields \cite{pgrassi}.} and the 
corresponding Faddeev-Popov terms are invariant under background gauge 
transformations and this leads to
supplementary Ward-Takahashi Identities (WTI)  (compare with \cite{bkg} and  \cite{msbkg}) 
which provide a useful tool for higher loop  
computations in gauge models and to keep under control some of the spurious anomalies 
generated by a non-symmetric renormalization scheme.  

In the source terms ${\cal L}^{S.T.}$ the BRST 
transformations of quantized fields $A^{a_S}_{\mu}, c_{a_S}, \phi_i, \psi_I$  
are coupled to their external sources 
$\gamma^{a_S}_{\mu}, \zeta_{a_S}, \gamma_i, \eta_I$ and the BRST transformations 
for the remaining fields  $A^{a_A}_{\mu}, c_{a_A}, \h{A}^{a}_{\mu}, \h{\phi}_i,  \bc_{a}, b_{a}$, 
can be immediately read from the STI in the appendix A. 

By taking the derivatives of the classical action $\gc$ with respect to 
the abelian ghost fields $c^{a_A}$, we immediately obtain 
\begin{eqnarray}
  \label{pre_aae}
  \fdu{\gc}{c}{a_A} 
  & = & \del^{a_{A}b} \partial^{2} \bc_{b} +  
        \left[ \bc_{b} \rho^{bc} (\h{\phi}+v)_{i} (et)^{c}_{ij} 
        (et)^{a_A}_{jk}(\phi+v)_{k} \right] + \nonumber \\
  && + \gam^{i} (et)^{a_A}_{ij} (\phi+v)^{j} + \b{\eta}_{I} (eT)^{a_A}_{IJ} \psi_{J} 
  + h.c. =  
        \nonumber \\ 
  & = & \del^{a_{A}b} \partial^{2} \bc_{b} +  
        (\h{\phi}+v)_{i} (et)^{a_A}_{ji} 
        \left[  \bc_{b} \rho^{bc} (et)^{c}_{ik} 
        (\phi+v)_{k} \right] + \nonumber \\
  && + \gam^{i} (et)^{a_A}_{ij} (\phi+v)^{j} + \b{\eta}_{I} (eT)^{a_A}_{IJ} \psi_{J} 
  + h.c. 
\end{eqnarray}
In the second line we have commuted the abelian generators in the  
representation for scalars. 
The composite operators which appear in the square brackets 
are coupled to the external fields $\Om_{i}$. In fact by 
differentiating $\gc$ with respect to them, we get:
\begin{eqnarray}
  \label{diffe}
   \fdu{\gc}{\Om}{i} =  \dfu{\Om}{i} \dms{\int \dx} {\cal L}^{\fp} = 
\bc^{a} \rho_{ab} t^{b}_{ij} (\phi+v)_{j}  
\end{eqnarray}
and comparing (\ref{pre_aae}) and (\ref{diffe}) we get the 
functional Abelian Anti-ghost Equation:
\begin{eqnarray}
  \label{aae}
  \fdu{\gc}{c}{{a_{A}}}-(et)^{a_{A}}_{kj} (\h{\phi}+v)_{k} 
\fdu{\gc}{\Om}{j}
= \partial^{2} \bc^{a_{A}} + \gam_{j}(et)^{a_{A}}_{jk}(\phi+v)_{k}
  + \b{\eta}_{I} (eT)^{a_A}_{IJ} \psi_{J} + h.c. 
\end{eqnarray}

In this form the AAE can be implemented at the quantum level ( 
by replacing the vertex functional $\g_0$ with $\g$) and one has only 
to check, by using the usual algebraic techniques (\cite{bec}, \cite{henri} and \cite{algeb}),  
that the eq. (\ref{aae}) does not get any 
quantum corrections which cannot be removed, order-by-order, 
by means of counterterms in the 
tree level action.  The only possible non-removable corrections to the r.h.s. of the AAE 
are the IR anomalies\footnote{I am grateful to E. Kraus 
to point out that in the complete on-shell scheme for the SM 
this situation occurs.}.  
The corresponding counterterms might cause IR 
divergences in the next order of perturbative expansion (some examples can be 
found in the papers by Bandelloni {\it et al.} \cite{radia} and by Clark 
{\it et al.} 
\cite{cla_si} and \cite{bard_si}).  
A carefully analysis on this point will be presented in the appendix B.

We would like to stress that only for the  
for abelian ghost fields $c^{a_A}$ an anti-ghost equation can be preserved to all 
order of perturbative expansion, since all  composite operators  present in the 
equation exist already at the tree level. The same strategy does not work for the non-abelian 
ghosts. 

The non-renormalization properties of the ghost fields can be proved also in the 
case of Landau gauge \cite{landau} 
(without background fields). 
In fact also in the Landau gauge an antighost equation can be derived. However in this 
context the composite operators, obtained by differentiating $\g$ with respect to the 
ghost fields, can be expressed in terms of functional derivatives of fields only if 
the antighost equation is integrated over Minkowski space-time.

Note that on r.h.s. of the AAE contains the couplings of the abelian ghost fields to
the BRST sources, scalars and fermions (according to the representation of abelian generators 
of the Lie algebra ${\cal G}$). 
If the AAE can be implemented to higher orders, it assures that 
these representations are stable against the radiative corrections. In fact 
by differentiating the AAE with respect to Fourier transforms of the external source 
$\widetilde{\bar{\eta}}_{I}(p)$ and the 
fermion field $\widetilde{\psi}_{J}(q)$  we get 
\bea\label{fe_st}
\dms{\frac{\tilde{\del}^2 \del \g}{ \del \widetilde{\bar{\eta}}_{I}(p) 
\widetilde{\psi}_{J}(-p) \del c^{a_A}(0)}}\left. \right|_{\Phi=0} - 
v_{k} (et)^{a_{A}}_{kj}  \dms{\frac{\tilde{\del}^2 \del \g}{ \del
\widetilde{\bar{\eta}}_{I}(p) 
\widetilde{\psi}_{J}(-p) \del \Om^{j}(0)}}\left. \right|_{\Phi=0} = - (eT)^{a_A}_{IJ}
\eea 

The second term of the equation, namely the Green's function 
with external $\Om^{j}(0)$ is superficial convergent by power counting. Thus this 
equation guarantees that the Green's function 
${\g}_{\bar{\eta}_{I} {\psi}_{J} c^{a_A}}$
is finite. 

In the same way by differentiating with respect to Fourier transforms of the external source 
$\widetilde{{\gamma}}_{i}(p)$ and the 
fermion field $\widetilde{\phi}_{j}(q)$ the AAE gives
\bea\label{sc_st}
\dms{\frac{\tilde{\del}^2 \del \g}{ \del \widetilde{{\gamma}}_{i}(p) 
\widetilde{\phi}_{j}(-p) \del c^{a_A}(0)}}\left. \right|_{\Phi=0} - 
v_{k} (et)^{a_{A}}_{kl} 
\dms{\frac{\tilde{\del}^2 \del \g}{ \del \widetilde{{\gamma}}_{i}(p) 
\widetilde{\phi}_{j}(-p) \del \Om^{l}(0)}}\left. \right|_{\Phi=0} = - (et)^{a_A}_{ij}
\eea 
that fixes the coupling of the ghost fields $c^{a_A}$ to the 
scalar fields. 

In order to impose 
suitable normalization conditions on the model ({\it e.g.} on-shell normalization 
conditions) an effective 
action (in the sense of Lowenstein and Zimmermann, \cite{low}, \cite{pig} and 
the reference therein) has to be used. In the effective action the fields and the free parameters 
are rescaled by means of finite renormalization factors (in the following we will denote 
those factors by ``Z''). However 
some 
attention must be paid to the linear (in the quantized fields) terms  
\beq\label{cla_lin}
\Delta^{a_A} = 
\partial^{2} \bc^{a_{A}} + \gam_{j}(et)^{a_{A}}_{jk}(\phi+v)_{k}
+ \b{\eta}_{I} (eT)^{a_A}_{IJ} \psi_{J} + h.c. 
\eeq
of the eq. (\ref{aae}). 

In fact by replacing the fields and the charges with 
the rescaled ones in 
the linear part we lose partially the informations contained there. 
One holds the information
on the functional structure of the local part 
of Green's functions with external ghost fields. However  this is not sufficient
to guarantee the stability for non-semi-simple gauge models. 
In this situation the AAE assures that the anomalies  
${\cal A}_{1}, {\cal A}_2$ described in the equations (\ref{ano_1}) and 
(\ref{ano_2})
never appear and therefore this implies that the anomaly terms as 
(\ref{cp_sm}) are not present in the SM without  discrete CP symmetry). However  
for the zero Faddeev-Popov charge 
sector we deduce the same structure by a simple power counting analysis 
where the terms $\g^{Inst}$ are allowed. 

What we really need is to impose the tree level charge structure.  
As an example in the SM  
we want to impose  that  only the abelian generator of the hyper-charge  
couples to the abelian gauge field $A^{\cal Y}_{\mu}$.
  
Henceforth we have to impose the functional equation for ghost 
field as in the classical approximation, that is with its linear terms 
$\Delta^{a_A}$, up to field and 
source ``Z'' factors and, more important, 
up to the ``Z'' factors of the non-vanishing entries of the 
charge tensor $e_{a_A b_A}$. 
For the SM, as will be explained in the following section, only the   
$Z_{g_1}$ for hyper-charge is allowed and this 
allows 
to fix the electric charge by the vertex photon-electron-positron in the 
Thompson limit. 
In general we have to impose that 
the vanishing entries of the charge tensor remain zero and 
the non-vanishing entries get their proper "Z" factors. 
Clearly this conditions could be imposed by hand on the 
Green's functions, however only the anti-ghost equation (or the 
QED-WTI as in the Kraus' work \cite{krau_ew})
assures that order-by-order these normalization conditions 
can be satisfied.

Finally, we 
would like to remind that already in \cite{band} and \cite{henri}, a 
similar functional equation is derived for the purpose of checking the 
couplings of the abelian ghost fields. 
In our approach  the AAE is a consequence of the quantization
in the `t Hooft-background gauge. In fact 
it expresses the background gauge invariance of the model. 
The  Ward-Takahashi Identities (WTI) associated to this invariance 
can be deduced by using 
${\cal E}^{a_A} \equiv \dfu{c}{{a_{A}}}-(et)^{a_{A}}_{jk} (\h{\phi}+v)_{k} 
\dfu{\Om}{j}$ 
and by considering the anti-commutator 
of the Slavnov-Taylor operator ${\cal S}_{\g}$ 
with the AAE.  We get:
\bea\label{commu}
&&
{\cal S}_{\g} \left( {\cal E}^{a_A}(\g) - \Delta^{a_A} \right) + 
 {\cal E}^{a_A} {\cal S}(\g) = \nonumber \\
&& =  - \de_{\mu} \fd{\g}{A}{{a_{A}}}{\mu} + (et)^{a_A}_{ji} \left[
  (\phi+v)_{j} \fdu{\g}{\phi}{i} + (\h{\phi}+v)_{j}
  \fdu{\g}{\h{\phi}}{i} + \Om_{j} \fdu{\g}{\Om}{i}+ \gam_{j}
  \fdu{\g}{\gam}{i} \right] + \\
&& \hspace{.5cm}+ (eT)^{a_A}_{JI} \left[
\bar{\psi}_{I} \fdu{\g}{\bar{\psi}}{J}  
+  \fdu{\g}{{\psi}}{I} {\psi}_{J}
+  \bar{\eta}_{I} \fdu{\g}{\bar{\eta}}{J}
+  \fdu{\g}{{\eta}}{I} {\eta}_{J} \right] - \de^2 b^{a_A} = 0 \nonumber
\eea   
Since these WTI follow from the STI and the AAE they 
provide the same informations on the couplings of the abelian gauge fields.

In the next section we will translate the AAE into the physical 
field variables to be useful for practical computations. 
Eq. (\ref{commu}) provides a further check of the integrability    
of the complete system of functional equations in the BFM quantization.

\section{Application to the Standard Model}

As is well known in the minimal SM the fermion content (quantum fields and
their corresponding BRST sources) is  
$$
\psi_I = 
\left\{ u^L_{\alp}, d^L_{\alp}, e^L_{\alp}, \nu^{L}_{\alp},  
u^R_{\alp}, d^R_{\alp}, e^R_{\alp} 
\right\} \hspace{2cm}
\eta_I = 
\left\{ \eta^{u,L}_{\alp}, \eta^{d,L}_{\alp}, \eta^{e,L}_{\alp}, \eta^{\nu,L}_{\alp},  
\eta^{u,R}_{\alp}, \eta^{d,R}_{\alp}, \eta^{e,R}_{\alp} 
\right\}
$$
where $\alp$ denotes the flavour number of the fermions and the 
superscript $L,R$ their chirality (The color index for quarks is 
omitted). The free massless Dirac action 
$ {\sum_{I}} {\int} \dx\b{\psi}^{I}    
\sla{\, \partial} \psi_{I}$ turns out to be invariant under the large 
global group  $U(21)$ corresponding to all  
rotations among fermion fields with different 
quantum numbers. This group is reduced by imposing the absence of any  
mixing between the flavour group $U(3)$ 
and the remaining unitary group $U(7)$. And then $U(7)$ is further reduced down to ${U(2)} \otimes 
{U(2)} \otimes {U(2)} \otimes {U(1)}$ by requiring the absence of rotations mixing 
fermions with opposite chirality and mixing quarks with 
leptons. Finally gauging the $SU(2) \otimes U(1)$ group and 
grouping together the doublets $ Q^L_{i,\alp} = \left( u^L_{\alp}, d^L_{\alp} \right), ~~~ 
L^L_{i,\alp} = \left( \nu^L_{\alp}, e^L_{\alp} \right)$, we 
are left with a residual  $U(1)^5 \otimes U(3)$ symmetry. 

By means of the the Yukawa terms 
\beq\label{yu_te}   
\dms{\int} \dx \left(   
Y^{l}_{\alp\bet}\b{L}^{i}_{L,\alp}(\Phi+v)_{i} e_{R,\bet} +    
Y^{u}_{\alp\bet}\b{Q}^{i}_{L,\alp}(\widetilde{\Phi}+\tilde{v})_{i} u_{R,\bet} +    
Y^{d}_{\alp\bet}\b{Q}_{L,\alp}(\Phi+v)_{i} d_{R,\bet} + h.c. \right)    
\eeq           
where $\Phi_i$ are the Higgs and would-be Goldstone fields in the 
fundamental representation of $SU(2)$  and $\widetilde{\Phi_{i}}= \left( 
i\tau^2 \Phi^* \right)_{i}$ is the 
charge conjugated of $\Phi_{i}$, the residual symmetry $U(1)^5 \otimes U(3)$ is finally 
broken to the abelian group\footnote{The Yukawa terms apparently break the residual symmetry 
in such way that only 
the hyper-charge, baryon and total lepton number are conserved, however by  two    
independent bi-unitary transformations  the matrices $Y^{l}_{\alp\bet}, Y^{d}_{\alp\bet}$ 
can be diagonalized and the individual lepton number are also conserved quantum numbers.}
\begin{eqnarray}\label{decompo}
\ber{cccccc} 
U(1)^5 =  & U(1) & \otimes &  U(1) & \otimes & U(1)^3 \\
& {\cal Y} && B && L_{\alp} ~~\alp=1,2,3
\eer
\end{eqnarray}
where ${\cal Y},B, L_{\alp} ~~\alp=1,2,3$ are respectively the 
conserved hyper-charge number, the baryon number, and  
the individual lepton numbers.  

To each residual conserved quantum number corresponds a 
conserved Noether current $j^{a_A}_{\mu} =    
i\dms{\sum_{I}} \b{\psi}_{I} T^{a_A}_{IJ}     
\gamma_{\mu} \psi_{I}$ where $T^{a_A}_{IJ}, ~~ a_A ={\cal Y},B, L_{\alp}$ 
are the generators of each    
$U(1)$ group of the decomposition (\ref{decompo}),  
explicitly given by the following expressions:
\begin{eqnarray} 
&&\hspace{-1cm} j^{B}_{\mu}  =  \sum_{\alp} \left( \b{Q}_{L,\alp} \gamma_{\mu} Q_{L,\alp} + 
 \b{u}_{R,\alp} \gamma_{\mu} u_{R,\alp} +  \b{d}_{R,\alp} \gamma_{\mu} d_{R,\alp}  
\right)  \nonumber  \\
&&\hspace{-1cm} j^{L_{\alp}}_{\mu} =  \left(
\b{L}_{L,\alp} \gamma_{\mu} L_{L,\alp} + \b{e}_{R,\alp} \gamma_{\mu} e_{R,\alp} \right) \\
&&\hspace{-1cm} j^{\cal Y}_{\mu} = \sum_{\alp} \left( \frac{1}{6} \b{Q}_{L,\alp} \gamma_{\mu} Q_{L,\alp} + 
\frac{2}{3} \b{u}_{R,\alp} \gamma_{\mu} u_{R,\alp} -  \frac{1}{3} 
\b{d}_{R,\alp} \gamma_{\mu} d_{R,\alp} - \frac{1}{2} \b{L}_{L,\alp} 
\gamma_{\mu} L_{L,\alp} - \b{e}_{R,\alp} \gamma_{\mu} e_{R,\alp} \right) \nonumber 
\end{eqnarray}

In the minimal SM only the hyper-charge current $j^{\cal Y}_{\mu}$
is coupled to the abelian gauge field $A^{\cal Y}_{\mu}$ and the 
charge tensor $e_{a_A b_A}$ is given by:
\bea\label{ch_te_sm}
e_{{\cal Y},{\cal Y}} = g_1, ~~~ e_{{\cal Y}, L_{\alp}} = 0, ~~~ e_{{\cal Y}, B} = 0, \nonumber \\
e_{L_{\bet}, L_{\alp}} = 0, ~~~ e_{B, L_{\alp}} = 0, ~~~ e_{B, B} = 0  
\eea
where the only non-vanishing entry $g_1$ is the hyper-charge gauge coupling. The 
BRST sources are given in the physical field representation by
\bea\label{cic}
\hspace{-1cm}
{\cal L}^{S.T.,\psi} & = & \sum_{\alp}  \left[ 
\b{\eta}^{e,L}_{\alp} 
\left[
\frac{ie}{\sqrt{2}s_W } c^{-} \nu_{L,\alp} - i\,e\, \left(
\left( \frac{1}{2 s_W c_W} - \frac{s_W }{c_W} \right) c_Z + c_{\gamma} \right)
e_{L,\alp}
\right] 
+ \right. \nonumber \\  
&& \left.
+
\b{\eta}^{\nu,L}_{\alp} 
\left[
\frac{ie}{\sqrt{2}s_W } c^{+} e_{L,\alp} +  \frac{i\,e }{2 s_W c_W} c_Z 
\nu_{L,\alp} \right]
+ \right. \nonumber \\  
&& \left.
+
\b{\eta}^{e,R}_{\alp} 
\left[- i\,e\, \left( c_{\gamma}
- \frac{s_W }{c_W} c_Z   \right) e_{R,\alp}
\right] 
+ \right. \nonumber \\  
&& \left.
+
\b{\eta}^{u,L}_{\alp} 
\left[
\frac{ie}{\sqrt{2} s_W } c^{+} d_{L,\alp} - i\,e\, \left(
\left( \frac{1}{2 s_W c_W} - \frac{2 s_W }{3 c_W} \right) c_Z +
\frac{2}{3} c_{\gamma} \right)
u_{L,\alp}
\right]
+ \right. \nonumber \\  
&& \left.
+
\b{\eta}^{d,L}_{\alp} 
\left[
\frac{ie}{\sqrt{2} s_W } c^{-} u_{L,\alp} - i\,e\, \left(
\left( \frac{1}{2 s_W c_W} + \frac{s_W }{3 c_W} \right) c_Z -
\frac{1}{3} c_{\gamma} \right)
d_{L,\alp}
\right]
+ \right. \nonumber \\  
&& \left.
+
\b{\eta}^{u,R}_{\alp} 
\left[- \frac{2\,i\,e}{3} \left( c_{\gamma}
- \frac{s_W }{c_W} c_Z   \right) u_{R,\alp}
\right] 
+ \right. \nonumber \\  
&& \left.
+
\b{\eta}^{d,R}_{\alp} 
\left[\frac{\,i\,e}{3} \left( c_{\gamma}
- \frac{s_W }{c_W} c_Z   \right) d_{R,\alp}
\right] 
+ h.c.  \right]
\eea 

Hence the terms of $\g^{Inst}_0$ (\ref{insta}) assume the form:
\beq\label{ins_sm}
\sum_{a_A} k^{a_A}  \dms{\int \dx}  
\left( j^{\mu}_{a_A} A^{\cal Y}_{\mu} +  
\bar{P}^{I}_{a_A}  \eta_I c^{\cal Y} + 
P^{I}_{a_A} \bar{\eta}_I c^{\cal Y} 
\right) 
\eeq
where $P^{I}_{a_A} = T^{a_A}_{IJ} \psi_{J}, \bar{P}^{I}_{a_A} = 
\bar{\psi} (T^{\dagger})^{a_A}_{IJ}$ and $c^{\cal Y}$ is the abelian ghost field 
related to the $c_Z, c_{\gamma}$ by means of Weinberg's rotation (about   
Weinberg's rotation  for ghost fields 
see the paper \cite{krau_ew} or the appendix B).       

Because of instability terms (\ref{ins_sm}) the couplings between the photon 
and the matter fields are modified. This can be easily seen 
by computing the modified electric charges of leptons
\begin{eqnarray}
\left\{
  \begin{array}{c}
\left[ Q^{e}, 
  \left(  
    \begin{array}{c}
     \nu_{L,\alp}   \\ e_{L,\alp}
    \end{array} 
  \right)
\right] 
=   
\left(
  \begin{array}{c}
  e \frac{k^{\cal Y} + k^{\alp}}{2\, g_1} \nu_{L,\alp} \\
  e \left( -1 + \frac{k^{\cal Y}+k^{\alp}}{2 \, g_1} \right) e_{L,\alp}  
  \end{array} 
\right) \\
\left[ Q^{e}, e_{R,\alp} \right] =  
e \left( -1 + \frac{k^{\cal Y}+2 k^{\alp}}{2 \, g_1} \right) e_{R,\alp}
 \end{array}
\right.
\end{eqnarray}
where $e$ denotes the electric charge of the electron. 
Note also that if $k^{\alp}\neq 0$ the electric charge of the left-handed 
and the right-handed part of the electron are different. 
In the same way one can derive the deviation from the n\"aive quark charges. 
Even more dramatically the neutrinos become electrically charged, or, equivalently,    
the Gell-Mann-Nishijima relation between $SU(2)$ isospin and    
hyper-charges is broken by  radiative corrections.    
Therefore the model is not stable under radiative corrections and new    
hard couplings for the Abelian ghost fields appear.    

As discussed in the previous section, the way out of this problem is to 
introduce and to respect (by means of the renormalization procedure) the AAE 
which provides the correct non-renormalization properties for the fermion representations  
of abelian factor $U(1)$ coupled to the gauge boson of the SM. 
 
In the present context it is useful to write the AAE in the terms of the physical fields
\bea\label{AAE_phy}
&& c_W \fdu{\g}{c}{A} - s_W \fdu{\g}{c}{Z} + 
\frac{i e}{2 c_W} \left( \hat{G}^+ \fdu{\g}{\Om}{+} -  
\hat{G}^- \fdu{\g}{\Om}{-} \right) - 
\frac{e}{2 c_W} \left( \hat{G}_0 \fdd{\g}{\Om}{H} -  
(\hat{H}+v) \fdd{\g}{\Om}{0} \right) = \nonumber \\ 
&& \hspace{.5cm} = \frac{e}{2 c_W} \left( \gamma_H G_0 - 
\gamma_0 (H+v)\right)  + \frac{i e}{2 c_W} \left(\gamma^+ G^- -  \gamma^- G^+ \right) + 
\\
&& \hspace{.5cm} +  \sum_{\alp} \left( \frac{1}{6} \b{\eta}^{Q,L}_{\alp} Q^{L}_{\alp} + 
\frac{2}{3} \b{\eta}^{u,R}_{\alp} u^{R}_{\alp} -  \frac{1}{3} 
\b{\eta}^{d,R}_{\alp} d^{R}_{\alp} - \frac{1}{2} \b{\eta}^{L,L}_{\alp} 
L^{L}_{\alp} - \b{\eta}^{e,R}_{\alp} e^{R}_{\alp} \right)  + h.c. + 
\nonumber \\  
&& \hspace{.5cm} + \left(s_W \de^2 \bc^{Z} - c_W  \de^2 \bc^{Z} \right)
\nonumber 
\eea
where $ G_0, H, G^{\pm}$ and  $\hat{G}_0, \hat{H}, \hat{G}^{\pm}$ are the 
would-be Goldstone bosons and their corresponding background partners. $\gamma_0, 
\gamma_H, \gamma^{\pm}$ are the BRST sources for the variations of the scalar fields 
and $\Om_0, \Om_H, \Om^{\pm}$ those of background scalar fields. 
In the present equation the Weak angle ($c_W \equiv 
cos \theta_W$) can either be considered as given by its measured value  
or alternatively as derived quantity being function of gauge bosons masses. 
In the latter case the Weak angle obtains radiative corrections. 

We also have to notice that the AAE does not depend on the `t~Hooft parameters and 
on the gauge fixing parameters. 

In the next section we will discuss how to compute the coefficients 
of the instability terms and we will verify that in the common dimensional 
regularization scheme 
(with the `t Hooft-Veltman prescription \cite{maiso} for $\gamma^5$) 
no divergent instability coefficients show up for the lepton electric charges. 

\section{Computation of the Instability coefficients.} 

The main problem in the computation of the coefficients $k_{h}$ of the 
instability terms (\ref{insta}) is to disentangle the non trivial contributions to the 
charge renormalizations 
from the wave function renormalization  $Z^{\psi}_{IJ}$, 
$Z^{\eta}_{IJ}$ and $Z^{C}_{ab}$ respectively for the fermions, their BRST sources 
and the  ghost fields
\footnote{$Z^{C}_{ab}$ also take into account the renormalization of 
mixing angles (such as the Weak angle $\theta_{W}$). For this point 
we refer to the recent work by E. Kraus (\cite{krau_ew}), where a detailed discussion on the 
renormalization ghost mixing angle  is presented} which have to be computed from the 
two points functions for fermion fields and the two points functions for ghost fields. 

As is clear from the functional structure of the instability 
terms  (\ref{insta}), their one loop coefficients $k^{(1)}_{h}$ can be equivalently 
derived from two different types of Green's functions 
\beq\label{2_k}   
\dms{\frac{\del^3 \g^{(1)}}{\del A^{a_A}_{\mu} \del \bar{\psi}_I \del \psi_J} } 
~~ {\rm or} ~~
\dms{\frac{\del^3 \g^{(1)}}{\del c^{a_A} \del \bar{\eta}_I \del \psi_J} } 
\eeq
since they are related by the STI (compare the
paper by Aoki {\it et al.} \cite{aoki} and the
references therein, where a discussion is given).
However in view of  their relevance for two loop calculations, it is instructive 
to describe how to extract the charge renormalizations from Green's functions 
with external ghost fields and BRST sources and to discuss their renormalization.   

To generate all possible counterterms (invariant and non-invariant ones) in the present 
sector is 
sufficient to rescale the quantum fields and sources by wave function renormalizations
\beq\label{w_q} 
\psi_{I} \longrightarrow Z^{\psi}_{IJ}\psi_{J}, ~~~
\eta_{I} \longrightarrow Z^{\eta}_{IJ}\eta_{J}, ~~~
c_{a} \longrightarrow Z^{C}_{ab} c_{b}
 \eeq
and to perform a charge renormalization in the following way
\beq\label{c_r}
e_{S} \longrightarrow Z^{e}_{S} e_{S}, ~~~~
e_{a_A,b_A}  \longrightarrow \Theta_{a_A,c_A} e_{c_A,d_A} \bar{\Theta}_{d_A,c_A}
\eeq 
with the constraint $\Theta_{a_A,c_A} = \bar{\Theta}_{c_A,a_A}$ due to the  
symmetry of charge matrix $e_{a_A,b_A}$. 
In the one loop approximation, expanding all 
wave function renormalizations and charge renormalizations in a 
power series $Z = 1 + \sum_{n=0}^{\infty} \hbar^{n} \del Z^{(n)}$,  
the local part of the Green's functions $
{\frac{\del^3 \g^{(1)}}{\del c^{a_A} \del \bar{\eta}_I \del \psi_J} } $ is expressed by
\bea\label{local} 
\mbox{\bf \Large Y}^{a_A}_{IJ} & \equiv & \lim_{p \rightarrow \infty} 
\dms{\frac{\del^3 \g^{(1)}}{\del c^{a_A} \del \widetilde{\bar{\eta}_I}(p) 
\widetilde{\del \psi_J}(-p)} }  
= \nonumber \\
&& =   \del Z^{(1),C}_{a_A, b_A} e_{b_A,c_A} T^{c_A}_{IJ} 
+ \del Z^{(1),C}_{a_A, b_S} e_{S} T^{b_S}_{IJ} 
+ \del \Theta^{(1)}_{a_A,c_A} e_{c_A,d_A} T^{d_A}_{IJ} + \nonumber \\
&& 
+\,  e_{a_A,b_A} \del \bar{\Theta}^{(1)}_{b_A,c_A} T^{c_A}_{IJ} 
+ \del Z^{(1),\eta}_{IK} e_{a_A,b_A} T^{b_A}_{KJ} 
+ e_{a_A,b_A} T^{c_A}_{IK} \del Z^{(1),\psi}_{KJ}  
\eea

Choosing the normalization for non-abelian generators ${\rm tr}\left[ t^{a_S} t^{b_S} \right] 
= C_{S} \delta^{a_S,b_S}$ and by using ${\rm tr}\left[ t^{a_S} t^{b_A} \right] 
= 0$ we can extract those terms which contain non-abelian generators:
\beq\label{e_an}
\hat{\mbox{\bf \Large Y}}^{a_A}_{IJ} = \mbox{\bf \Large Y}^{a_A}_{IJ} - \frac{1}{C_S} 
{\rm tr}\left[ t^{a_S} \mbox{\bf \Large Y}^{a_A}\right]_{IK} t^{a_S}_{KJ}
\eeq
We subtract the pieces proportional to the 
w.f.r. of the fermion fields, BRST sources\footnote{By means of the 
STI, the w.f.r. of the BRST sources, namely $ Z^{\eta}_{IJ}$, are related to the w.f.r. 
$Z^{\psi}_{KJ}$
of fermion fields by 
$\sum_{K} Z^{\eta}_{IK} Z^{\psi}_{KJ} = const~\times \delta_{IJ}$, where the constant  
depends on the conventions for the w.f.r. of the anti-ghost fields $\bar{c}^{a}$. }  
and ghost fields
\beq\label{e_an_1}
\widetilde{\hat{\mbox{\bf \Large Y}}}^{a_A}_{IJ} =  
\del \Theta^{(1)}_{a_A,c_A} e_{c_A,d_A} T^{d_A}_{IJ}
+\,  e_{a_A,b_A} \del \bar{\Theta}^{(1)}_{b_A,c_A} T^{c_A}_{IJ} 
\eeq

Since the abelian generator $T^{d_A}_{IJ}$ can be simultaneously 
diagonalized\footnote{The generators $T^{a_A}_{IJ}$ commute with 
the  $T^{a_S}_{IJ}$ and since the simple factor for the SM, namely $SU(2)$, 
act differently on the left and right handed fermions, the generators $T^{a_A}_{IJ}$ 
cannot mix left and right fermions and they can be diagonalized completely.
} 
by means of unitary transformations, (denoting  
their eigenvalues by $\lambda^{a_A}_{I}$ ), equation (\ref{e_an_1}) 
becomes 
\beq\label{e_an_2}
\widetilde{\hat{\mbox{\bf \Large Y}}}^{a_A}_{II} =  
\left( \del \Theta^{(1)}_{a_A,c_A} e_{c_A,d_A} +  
e_{a_A,b_A} \del \bar{\Theta}^{(1)}_{b_A,d_A} \right) \lambda^{d_A}_{I} 
\eeq

In the SM, the index $I$, which labels the fermion fields, 
compactly expresses all fermion quantum numbers such as the flavour, the 
color, the type, the chirality and the isospin. Furthermore the 
abelian index ${a_A}$ corresponds to the 
hyper-charge (the only gauged abelian generator) and the indices $c_A,b_A,d_A$ 
correspond to the five generators of the 
hyper-charge, the lepton numbers and the baryon number. In this 
context the charge matrix  $e_{a_A,b_A}$ is given in the eq. (\ref{ch_te_sm}). 

Now the charge renormalization $ \del \Theta^{(1)}_{a_A,c_A}, 
\del \bar{\Theta}^{(1)}_{a_A,c_A}$ are given as solution of the system:
\bea\label{sys_ch_re}
\left\{ \ber{ccc}
g_1 \left( 
2 \del \Theta^{(1)}_{{\cal Y},{\cal Y}} \left( - \frac{1}{2} \right) 
+ \del \Theta^{(1)}_{L_{\alp},{\cal Y}} \right)  
& = & \widetilde{\hat{\bf  Y}}^{{\cal Y}}_{e^L_{\alp},e^L_{\alp}} \\
g_1 \left( 
2 \del \Theta^{(1)}_{{\cal Y},{\cal Y}}  
+ \del \Theta^{(1)}_{L_{\alp},{\cal Y}} \right)  
& = & \widetilde{\hat{\bf Y}}^{{\cal Y}}_{e^R_{\alp},e^R_{\alp}}  \\
g_1 \left( 
2 \del \Theta^{(1)}_{{\cal Y},{\cal Y}} \left( \frac{1}{6} \right) 
+ \del \Theta^{(1)}_{B,{\cal Y}} \right)
& = & \widetilde{\hat{\bf Y}}^{{\cal Y}}_{Q^L_{\alp},Q^L_{\alp}}   \\
 g_1 \left( 
2 \del \Theta^{(1)}_{{\cal Y},{\cal Y}} \left( \frac{2}{3} \right) 
+ \del \Theta^{(1)}_{B,{\cal Y}} \right)  
& = & \widetilde{\hat{\bf Y}}^{{\cal Y}}_{u^R_{\alp},u^R_{\alp}} \\
g_1 \left( 
2 \del \Theta^{(1)}_{{\cal Y},{\cal Y}} \left( - \frac{1}{3} \right) 
+ \del \Theta^{(1)}_{B,{\cal Y}} \right)   
& = & \widetilde{\hat{\bf Y}}^{{\cal Y}}_{d^R_{\alp},d^R_{\alp}} 
\eer
\right.
\eea

The renormalization constants $
\del \Theta^{(1)}_{B,{\cal Y}}, \del \Theta^{(1)}_{L_{\alp},{\cal Y}}$ which gives the 
coefficients of the deformations of the fermion representation (\ref{insta}), 
are equal to zero as has be shown in the paper by Aoki {\it 
et al.} \cite{aoki}.   

In particular it is very easy to see that in the computation of the instability terms 
for the lepton charges the divergent contribution in the Green's function 
\beq\label{one_loop_coe}
\left( \frac{\del \g^{(1)}}{\del \b{\eta}^{\nu}_{\alp} \nu_{\bet} c_{A}}
\right)_{DIV} = - \dms{\frac{ e^2 M_W}{8 \pi^2 \sqrt{M^2_Z-M^2_W}}}
\xi (1-\gamma^5) C_{UV}
\eeq
where $C_{UV}$ is given in \cite{aoki}, is re-absorbed by means of the  
w.f.r. $Z^{1/2}_{ZA}$, an element of the 
w.f.r. matrix of the 
$\gamma$-Z gauge bosons. Thus we can reabsorb this 
divergent contribution in a renormalization of the Weak angle $\theta_{W}$. 

This is possible because the coefficient of the divergent term is 
independent of the fermion type (quark or lepton) and of their 
couplings or masses. Thus in the present situation no new term 
comes up to destroy the stability of the fermion representation and 
modifying the electric charge of the neutrino. This is due to the 
dimensional regularization and to the absence of fermion loops in the computation 
of $\mbox{\bf Y}^{a_A}_{IJ}$ at one loop. 

In a forthcoming work a complete discussion on the renormalization of the 
Green's functions with BRST source terms will be presented. 

\section{Appendix A: Conventions and Slavnov-Taylor identities}
\label{appa}

The field content is specified by the quantized gauge vectors $A^{a}_{\mu}$, 
their background partners $\h{A}^{a}_{\mu}$, 
the scalars $\phi_{i}$, the background scalars $\h{\phi}_{i}$, 
the fermions $\psi_I = \left\{\psi_L, \psi_R \right\}$, 
the Faddeev-Popov ghosts $c^{a}, \bc^{a}$, the Nakanishi-Lautrup multipliers $b^{a}$, 
the BRST sources $\gamma^{a}_{\mu}, \gamma_i, \eta_I, \zeta^{a}$ for 
quantized fields and the external fields $\Om^{a}_{\mu}, \Om_i$. 
The component of the fields $A^{a}_{\mu}, c^{a}, \bar{C}^{a}, b^{a}$  
are identified by the index $a$ labeling a basis of the Lie algebra ${\cal G}$ of the 
gauge group; the fields $\h{A}^{a_S}_{\mu}, \gamma^{a_S}_{\mu}, \zeta^{a_S}, \Om^{a_S}_{\mu}$ 
are restricted to the semi-simple factor  ${\cal G}_{S}$ of  ${\cal G}$ while 
$\phi_{i},\h{\phi}_{i}, \gamma_i,\Om_i; \psi_I, \eta_I $ define two different representation 
spaces for ${\cal G}$. The fields are also characterized by a conserved Faddeev-Popov charge 
which is: $0$ for  $A^{a}_{\mu}, \h{A}^{a}_{\mu},  \phi_{i}, \h{\phi}_{i}, 
\psi_I, b^{a}$, $-1$ for $\bar{C}^{a}, \gamma^{a_S}_{\mu}, \gamma_i, \eta_I$, $-2$ 
for $\zeta^{a_S}$ and $+1$ for $c^{a}, \Om^{a_S}_{\mu}, \Om_i$. 

In order to describe this general model it is useful to introduce the charge matrices 
involved in the coupling of the gauge fields $A^{a}_{\mu}$. First of all we specify the 
symmetric, positive definite charge matrix $e_{ab}$ on the adjoint representation of 
the algebra  ${\cal G}$. Clearly  
$e_{ab}$ has no elements connecting the semi-simple factor  ${\cal G}_{S}$ to 
the abelian one  ${\cal G}_{A}$. Furthermore the restriction of $e_{ab}$ 
to each simple component is proportional to the Killing form, and, in a basis 
where the latter is diagonal, we have $e_{a_S b_S}= e_S \del_{a_S b_S}$ and 
$ e_S$ is identified with the simple charges. Concerning the restriction of $e_{ab}$ 
to the abelian factor ${\cal G}_{A}$ the only requirement is the symmetry and the positive 
definiteness. 

The infinitesimal generators $t^{a}, T^{a}$ of the 
gauge group in the scalar and fermion representations obey 
\bea\label{repre}
&& (t^{a})^{T} = - t^{a}, ~~~~ (T^{a})^{\dagger} = - T^{a}  \\
&& \left[ t^{a},t^{b} \right] = f^{abc} t^{c} , ~~~~ \left[ T^{a},T^{b} \right] = f^{abc} T^{c}
\eea
where $t^{a}$ are real and $f^{abc}$ are structure constants of   ${\cal G}$, 
with $f^{abc_A} = 0$. 
The couplings 
of the gauge fields  $A^{a}_{\mu}$ can now be expressed in terms of the tensors 
\beq\label{cou_gau} 
(ef)^{a_S b_S c_S} = e_{S} f^{a_S b_S c_S}
\eeq
for the simple factors ${\cal G}_{S}$ and the couplings with scalar and fermion fields 
are given by:  
\beq\label{cou_sca}
(et)^{a} = e_{a b} t^{b},   ~~~~ (eT)^{a} = e_{a b} T^{b}
\eeq 

Since we are not interested in a complete and detailed discussion  on the 
renormalization of gauge models with background fields (compare 
the papers by \cite{pgrassi} and \cite{msgras}) we recall only some useful 
ingredients. 

In the main text the relevance of the the particular 
feature of 't Hooft-background gauge fixing and the BRST transformations  
of background fields $\h{A}^{a}_{\mu}, \h{\phi}_{i}$ are discussed. In the 
present context  
the corresponding Slavnov-Taylor identities for the 
vertex generating functional $\g$ are expressed by 
\bea\label{st}\ber{l}    
\oop{.3cm}    
{\cal S}(\g) = \dms{\int} \dx \left[    
\dms{\fd{\g}{\gamma}{a_S}{\mu}} \dms{\fd{\g}{A}{a_S}{\mu}} + 
\de_{\mu}c^{a_A} \dms{\fd{\g}{A}{a_A}{\mu}} +     
\dms{\fdd{\g}{\zeta}{a_S}} \dms{\fdu{\g}{c}{a_S}} +     
\dms{\fdd{\g}{\gamma}{i}} \dms{\fdu{\g}{\phi}{i}} +     
\dms{\fdd{\g}{\b{\eta}}{I}} \dms{\fdu{\g}{\psi}{I}} +     
 \right. \\    
\left. \hspace{2.5cm} +    
\dms{\fdd{\g}{\b{\psi}}{I}} \dms{\fdu{\g}{\eta}{I}} +     
b_{a} \mathfrak{G}^{ab} \dms{\fdd{\g}{\bar{C}}{b}} +     
\Om^{a_S}_{\mu} \dms{\fd{\g}{\h{A}}{a_S}{\mu}} +     
\Om_{i} \dms{\fdu{\g}{\h{\phi}}{i}} \right] = 0     
\eer \eea     
where the linear terms, proportional to the external fields $\Om^{a_S}_{\mu}, \Om_i$, 
control the background field dependence of the model. The general symmetric and invertible matrix
$\mathfrak{G}^{ab}$ 
was introduced in the papers by Becchi {\it et al.} 
\cite{bec}, \cite{henri}. It can be used to get rid of the possible IR problems 
in the on-shell renormalization scheme for the SM 
as pointed out by E.Kraus \cite{krau_ew}. However we can 
introduce the ``rotated'' anti-ghost fields $\bc^{a} = 
(\mathfrak{G}^{-1})^{ab}\bar{C}_{b}$ 
in order to simplify our derivations and we will use this matrix 
discussing the normalization conditions for the ghost fields in the appendix B. 

Besides to the Slavnov-Taylor identity, the Abelian Anti-ghost equation
(described in the text) and 
the Nakanishi-Lautrup equation
\bea\label{nl}
\fdu{\g}{b}{a}_{(x)} = \dms{\int \dy} \dfu{b}{a}_{(x)} {\cal L}^{g.f.}(y), 
\eea
as well as the auxiliary Faddeev-Popov 
equations 
\bea\label{fp}
&& \fdu{\g}{\bc}{b} +
    \del^{aa_{S}} \nabh ^{a_{S}b_{S}}_{\mu}
    \fd{\g }{\gam}{{b_{S}}}{\mu} + \rho^{ab} (\h{\phi}+v)_{i}
    (et)^{b}_{ij} \fdu{\g}{\gam}{j} = \nonumber \\
&& \hspace{1cm}
- \del^{aa_{A}} \de^{2} c_{a_{A}} -
    \del^{aa_{S}} \nabla_{\mu}^{a_{S}b_{S}} \Om^{b_{S}}_{\mu} -
    \rho^{ab} \Om_{i} (et)^{b}_{ij} (\phi+v)_{j} 
\eea
are useful analyzing the renormalization of the gauge model. 
The constant vector $v_{i}$ is the usual shift 
of the scalar fields, due to the 
spontaneous symmetry breaking. 
The constant matrix $\rho^{ab}$ are the 't Hooft parameters  
and  the covariant derivatives are defined by 
\bea\label{co_de} \ber{l}
\nabla^{a_S b_S}_{\mu} \Om^{b_S}_{\nu} = \partial_{\mu} \Om^{a_S}_{\nu} - 
(ef)^ {a_S b_S c_S} A^{b_S}_{\mu} \Om^{c_S}_{\nu} \nonumber \\
\nabh^{kl}_{\mu} \phi_{l} = \partial_{\mu} \phi^{k} - 
(et)^{ak}_{l} \h{A}^{a}_{\mu} \phi_{l}. \nonumber 
\eer \eea 
Notice that the Faddeev-Popov equations do 
not provide any information about the fermion representations.   
Furthermore the Faddeev-Popov eq. (\ref{fp}) does not contain more
information than the STI and the Nakanishi-Lautrup equation (\ref{nl}).
In fact the eq. (\ref{fp}) is obtained as the commutator of the
Slavnov-Taylor operator ${\cal S}$ and $\dfu{b}{a}$.

\section{Appendix B: Renormalization of the Faddeev-Popov equations and of the AAE}

In this appendix we discuss briefly the renormalization of the functional
equations for the ghost and anti-ghost fields. Although in the main 
text we have stressed the relevance of the AAE, here we use both
the Faddeev-Popov equation and the AAE in order to compute the necessary 
ghost dependent counterterms.

The present discussion follows essentially the lines of the proof of absence of
anomalies for the Faddeev-Popov equation and AAE in  \cite{henri} and 
in the paper  by T.Clark \cite{cla}. 
However in \cite{henri} the IR problems are not taken into account and in \cite{cla} 
only the Georgi-Glashow model is discussed. We will show
that in the presence of massless ghost fields as in the SM some
care has to be taken  to avoid IR problems and, for
a completely general model, non-removable anomalies could spoil the
functional identities.

To study the possible anomalous terms we
use to well established algebraic renormalization technique. 
To deal with massless and massive field the Feynman integrals
are treated by means of the Lowenstein and Zimmermann \cite{low} (BPHZL) subtraction
scheme.

According to \cite{low} the UV degree $d_{UV}$ and IR degree $d_{IR}$ are
respectively 1 for massless bosonic fields(except ghost fields),
$\frac{3}{2}$ for massless fermionic fields, $2$ for massive fields.
To avoid IR divergences we assign
$d_{UV} \bar{\omega}^{a} = 2, d_{UV} {\omega^{a}} = 0,
d_{IR} \bar{\omega}^{a} = 3$ and $d_{IR} {\omega^{a}} = 1$ for the massive
ghost fields $\omega^{a},  \bar{\omega}^{a}$ and 
$d_{UV} \bar{\chi}^{a} = 2, d_{UV} {\chi^{a}} = 0,
d_{IR} \bar{\chi}^{a} = 2$ and $d_{IR} {\chi^{a}} = 0$ for the massless
ghost fields $\chi^{a},  \bar{\chi}^{a}$. 

By denoting $\rho_{IR}$ as the IR subtraction degree,  $\delta_{UV}$ the UV
degree,  $N^{\rho_{IR}}_{\delta_{UV}}$ for normal products  and
introducing the functional operator
$\bar{\cal E}^{a} \equiv \dfu{\bc}{b} +
    \del^{aa_{S}} \nabh ^{a_{S}b_{S}}_{\mu}
    \dfa{\gam}{{b_{S}}}{\mu} + \rho^{ab} (\h{\phi}+v)_{i}
    (et)^{b}_{ij} \dfu{\gam}{j}$, the Faddeev-Popov equation and the AAE
at higher order are given by
\bea\label{fp_sy}
&& \bar{\cal E}^{a} \g = \bar{\Delta}^{a} + \left[\bar{Q}^{a} \cdot \g \right]^1_2 
\\ \label{aae_sy}
&& {\cal E}^{a_A} \g = {\Delta}^{a_A} + \left[ {Q}^{a_A} \cdot \g \right]^3_4 
\eea
where $\bar{\Delta}^{a}$ is the left hand side of eq. (\re{fp}) and ${\Delta}^{a_A}$ 
is given by (\ref{cla_lin}). 
Recursively assuming that the lower order breaking terms
of the eqs. (\ref{fp_sy})-(\ref{aae_sy}) up to order $\hbar^{n-1}$
are compensated by means of counterterms we deduce
\bea\label{fp_sy_q_0}
&& \left[\bar{Q}^{a} \cdot \g \right]^1_2 = \hbar^n \bar{Q}^{a} + O(\hbar^{n+1} \bar{Q}) 
\\ \label{aae_sy_q_0}
&& \left[{Q}^{a_A} \cdot \g \right]^3_4 = \hbar^n {Q}^{a_A} + O(\hbar^{n+1} {Q}) 
\eea
where $\bar{Q}^{a}, {Q}^{a_A}$ are local polynomials in terms of fields and
their derivatives with charge +1 and -1 respectively. 
By UV and IR power counting, by covariance and by Faddeev-Popov charges, the 
possible candidates for $\bar{Q}^{a}, {Q}^{a_A}$ are 
\bea\label{fp_sy_q}
&& \bar{Q}^{a} = \bar{X}^{a b}_1 c_b + \bar{X}^{a[b c]d}_2 c_b c_c \bc_d +
\bar{X}^{a i}_3 \Om_i + \bar{X}^{a b_S}_{\mu,4} \Om_{b_S}^{\mu} 
\\ \label{aae_sy_q}
&& {Q}^{a_A} = X^{a_A b}_1 \bc_b + X^{a_A [b c]d}_2 \bc_b \bc_c c_d +
X^{a_A i}_3 \gamma_i + X^{a_A b_S}_{\mu,4} \gamma_{b_S}^{\mu} + 
X^{a_A I}_{\mu,5} \bar{\eta}_I + h.c. 
\eea
where $\bar{X}^{a b}_1, \bar{X}^{a i}_3, \bar{X}^{a b_S}_{\mu,4}$,
${X}^{a_A b}_1, {X}^{a_A i}_3, {X}^{a_A b_S}_{\mu,4}, X^{a_A I}_{\mu,5}$ 
are polynomials of quantum fields while the constant coefficients 
$\bar{X}^{a[b c]d}$ are totally antisymmetric tensors with respect to the algebra ${\cal G}$.
This follows from the consistency conditions:
\bea\label{con_co}
&& \bar{\cal E}^{a}_{(x)} \bar{\cal E}^{b}_{(y)} +
\bar{\cal E}^{b}_{(y)} \bar{\cal E}^{a}_{(x)}  = 0 \nonumber \\
&& {\cal E}^{a_A}_{(x)}  {\cal E}^{b_A}_{(y)}  +
{\cal E}^{b_A}_{(y)}  {\cal E}^{a_A}_{(x)} = 0 \nonumber \\
&& {\cal E}^{a_A}_{(x)}  \bar{\cal E}^{b}_{(y)}  +
\bar{\cal E}^{b}_{(y)}  {\cal E}^{a_A}_{(x)} = 0
\eea 
which imply relations among the coefficients of $\bar{Q}^{a},{Q}^{a_A}$.

As is well known (see \cite{henri} and
\cite{pig} for further details) the breaking terms (\ref{fp_sy_q})-(\ref{aae_sy_q}) can be
removed by means of counterterms and no anomaly appears for equations
(\ref{fp_sy})-(\ref{aae_sy}). But, although from an algebraic point
of view there are local counterterms which cancel the apparent breaking terms,
we have to be sure that those counterterms do not introduce any IR
divergences. To this purpose we have to check the structure of the lower
dimensional terms in the explicit decomposition  (\ref{fp_sy_q})-(\ref{aae_sy_q}),
that is $\bar{X}^{a b}_1, {X}^{a_A b}_1$ and check if the corresponding counter
terms could give IR problems.
It is easy to see that the only dangerous candidates are
\beq\label{da_cou}
{\cal L}^{c.t.}_{\rho_{IR} \leq 3}(x) = \bar{\omega}^{a} K_{a,b}^1 \chi^{b} +
\bar{\chi}^{a} K_{a,b}^2 \omega^{b} + \bar{\chi}^{a} K_{a,b}^3(\phi,\h{\phi}) 
\chi^{b} 
\eeq
with IR degree $\rho_{IR} \leq 3$. Only $K_{a,b}^3(\phi,\h{\phi})$ could
depend on the scalars if they are massless, otherwise the coefficients are
constant and represent mass terms for the massless ghost fields $\chi,\bar{\chi}$.
However the last term of the ${\cal L}^{c.t.}_{\rho_{IR} \leq 3}$ 
is not necessary; in fact
IR power counting implies that
\bea
\label{fp_sy_q_1}
\bar{Q}^{a}_{\rho_{IR} \leq 1} & = & \bar{X}^{a b}_{1,con} \omega_b \\
\label{aae_sy_q_1}
{Q}^{a_A}_{\rho_{IR} \leq 3} & =  & X^{a_A b}_{1,con} \bar{\omega}_b
\eea
where $\bar{X}^{a b}_{1,con}, X^{a_A b}_{1,con}$ are the field independent
parts of the polynomials $\bar{X}^{a b}_{1}, X^{a_A b}_{1}$ and the
index $b$ runs only over the indices of massive ghost fields $\omega^a$.
Furthermore the consistency conditions (\ref{con_co}) impose the
constraint $\bar{X}^{a a_A}_{1,con} + X^{a_A a}_{1,con} = 0$,
reducing  the only free coefficients  to $\bar{X}^{a b}_{1,con}$ . 

In the case $\bar{X}^{a b}_{1,con}\neq 0$ the corresponding
counterterms (\ref{da_cou}) cannot be introduced in the tree level
action. However another solution can be found. We can use
the matrix $\mathfrak{G}^{ab}$ introduced above in order to remove the anomaly terms
(\ref{fp_sy_q_1})-(\ref{aae_sy_q_1}), or equivalently, to fix the normalization
conditions
\beq\label{no_con}
\g_{\bar{\omega}^a \chi^b}(p^2 = 0) = 0, \hspace{.5cm}
\g_{\bar{\chi}^a \omega^b}(p^2 = 0) = 0, \hspace{.5cm}
\g_{\bar{\chi}^a \chi^b}(p^2 = 0) = 0
\eeq
assuring the correct normalization properties of massless ghost fields.
As is well known the IR problems arise when radiative corrections
mix massive and massless fields. Therefore the anomalies in the
functional equations for the ghost fields can be removed by rotating
the anti-ghost fields. Finally we would like to stress that the
coefficients $\bar{X}^{a b}_{1,con}$ computed within BPHZL
scheme are zero and the normalization conditions
(\ref{no_con}) are automatically satisfied. On the other side
the choice of other normalization conditions for physical fields
(as for the Standard Model with on-shell normalization conditions)
might spoil eqs. (\ref{no_con}) and spurious anomalies as 
(\ref{fp_sy_q})-(\ref{aae_sy_q}) might appear.
This result is in complete agreement with results obtained by E.Kraus in
\cite{krau_ew}. 

\vspace{.5cm}    
{\bf \Large Acknowledgments}
\vspace{.2cm}

I am grateful to R.Ferrari and D.Maison for their assistance and helpful 
discussions. I am also grateful to E.Kraus for discussions and for sending 
me her paper on Renormalization of the Electroweak Model before publication. 

%
% references
%
\def\ap#1#2#3{Ann.\ Phys.\ (NY) #1 (19#3) #2}
\def\ar#1#2#3{Ann.\ Rev.\ Nucl.\ Part.\ Sci.\ #1 (19#3) #2}
\def\cpc#1#2#3{Computer Phys.\ Comm.\ #1 (19#3) #2}
\def\jmp#1#2#3{J.\ Math.\ Phys.\ #1 (19#3) #2}
\def\ib#1#2#3{ibid.\ #1 (19#3) #2}
\def\np#1#2#3{Nucl.\ Phys.\ B#1 (19#3) #2}
\def\pl#1#2#3{Phys.\ Lett.\ #1B (19#3) #2}
\def\pr#1#2#3{Phys.\ Rev.\ D #1 (19#3) #2}
\def\prb#1#2#3{Phys.\ Rev.\ B #1 (19#3) #2}
\def\prep#1#2#3{Phys.\ Rep.\ #1 (19#3) #2}
\def\prl#1#2#3{Phys.\ Rev.\ Lett.\ #1 (19#3) #2}
\def\rmp#1#2#3{Rev.\ Mod.\ Phys.\ #1 (19#3) #2}
\def\sj#1#2#3{Sov.\ J.\ Nucl.\ Phys.\ #1 (19#3) #2}
\def\zp#1#2#3{Zeit.\ Phys.\ C#1 (19#3) #2}
\def\cmp#1#2#3{Comm.\ Math.\ Phys.\ #1 (19#3) #2}
\def\cmp#1#2#3{Comm.\ Math.\ Phys.\ #1 (19#3) #2}
\def\nc#1#2#3{Il Nuovo Cimento #1A (19#3) #2}
\def\lnc#1#2#3{Lettere al Nuovo Cimento #1 (19#3) #2}
\def\apa#1#2#3{Acta Phy. Austriaca #1 (19#3) #2}

\eject 

\begin{thebibliography}{99}
                                                               
\bibitem{bec}
        C.Becchi, A.Rouet and R.Stora, \cmp{42}{127,141}{75},  
        \ap{98}{287,321}{76}; \\
        C.Becchi, A.Rouet and R.Stora, {\it Gauge field models} and {\it
        Renormalizable models with broken symmetries} in Renormalization 
        Theory, Edited by G.Velo and A.S.Wightman, D. Reidel, Pub. Co.; 
        C.Becchi, {\it Lectures on Renormalization of 
        Gauge Theories} in Relativity, group and topology II, Les Houches 1983
\bibitem{algeb}
        S.P.Sorella, \cmp{157}{231,243}{93}; 
        O.Piguet and S.P.Sorella \np{381}{373,393}{92};
        O.Piguet and A.Rouet, {\it Symmetries in Perturbative Quantum Field 
        Theory} \prep{76}{1,77}{81}; 
        G.Bandelloni, \jmp{28}{2775}{87};  
       J.A.Dixon, Cohomology and renormalization of gauge theories, 
        Imperial College preprint-1977; 
        L.Baulieu, \np{241}{557,588}{84}; 
        Z.Bern and A.G.Morgan, \pr{49}{6155, 6163}{94}; 
\bibitem{pig}
        O. Piguet and S.P. Sorella, {\it Algebraic Renormalization} 
        Lecture Notes in Physics Monographs m28, Springer-Verlag, 1995, 
        Berlin; 
\bibitem{CP}
        Kobayashi, M. and Maskawa, M. Prog. Theor. Phys. 43 (1979) 484;
\bibitem{henri}
        G.Bandelloni, C.Becchi, A.Blasi and R.Collina, 
        {\it Ann. Inst. Henry Poincar\'e} XXVIII, 3(1978)225,285; 
        C.Becchi, A.Rouet and R.Stora, {\it Renormalizable Theories with 
        Symmetry Breaking } in Field Theory, Quantization and Statistical 
        Physics, Ed. E. Tirapegui.
\bibitem{band}
       G.Bandelloni, C.Becchi, A. Blasi, R.Collina, \cmp{71}{239}{80};
\bibitem{henn}
        G.Barnich, F. Brandt,  
        M. Henneaux, Commun.Math.Phys.174:93-116,1995 {\bf hep-th/9405194},  
        Commun.Math.Phys.174:57-92,1995 {\bf hep-th/9405109}; 
        G.Barnich, F. Brandt,  
        M. Henneaux, Phys.Lett.B346:81-86,1995 {\bf hep-th/9411202} and the 
        references 
        therein ; G. Barnich , M. Henneaux  
        Phys.Rev.Lett.72:1588-1591,1994 {\bf hep-th/9312206}; 
\bibitem{krau_ew}
        E.Kraus, {\bf hep-th/9709154} to appear in Ann. Phys.;
\bibitem{bkg} 
        J.B.Zuber and H.Kluberg-Stern, \pr{12}{467,481}{75}, 
        \pr{12}{482,488}{75}; 
        G.t'Hooft \np{33}{436,450}{71}; 
        L.F.Abbott \np{185}{189,203}{81},
        Acta Phys.Pol.12(1982); 
        S.Ichimose and M.Omote \np{203}{221}{82}; 
        D.M.Capper and A.MacLean \np{203}{413}{82}; 
        D.G.Boulware, \pr{12}{389}{81};
\bibitem{msbkg}
        A.Denner, G.Weiglein and S.Dittmaier, \pl{333}{420,426}{94},
        {\it Nucl. Phys.}{\bf B}(Proceedings Supplements) 37B (1994) 87, 
        \np{440}{95}{95}; 
\bibitem{ren_mode}
        S.Weinberg, \np{469}{473}{96}; 
\bibitem{kugo}
        N.Nakanishi and I Ojima, {\it Covariant Operator Formalism of 
        Gauge Theories and Quantum Gravity } World Scientific Lecture Notes 
        in Physics Vol. 27, Ed. World Scientific, 1990. and the 
        reference therein. 
\bibitem{acci}
        S.Weinberg, {\it The Quantum Theory of Fields} Vol. II, 
        Cambridge University Press, 1996; T. Cheng and L.Li, {\it Gauge 
        Theory of elementary particle physics} Oxford University 
        Press , N.Y., 1984;
\bibitem{wess}
        J.Wess and B.Zumino, \pl{37}{95}{71}; 
\bibitem{abj}
        S.L.Adler, \pr{117}{2426}{69},
        J.S.Bell, R.Jackiw, \nc{60}{47}{69},              
        J.S.Bell, R.Jackiw, \nc{60}{47}{69},              
        J.Schwinger, \pr{82}{664}{51},
        W.A.Bardeen \pr{184}{1848}{69};
\bibitem{medrano}
        J. Dixon and M.Medrano, \pr{22}{429}{80}; 
\bibitem{wein}
        S.Weinberg, \pr{118}{838}{60};
\bibitem{fuji}
         K.Fujikawa, \pr{7}{393}{73};
\bibitem{appl_fuji}
         N.G. Deshpand and M. Nazerimonfared, \np{213}{390}{83}; 
\bibitem{pgrassi}
       P.A.Grassi, \np{462}{524}{96};
\bibitem{radia}
        G.Bandelloni, C.Becchi, A. Blasi, R.Collina, \cmp{67}{78}{147-178};
\bibitem{cla_si}
        T.E.Clark, O. Piguet and K.Sibold, \np{119}{292}{77}; 
\bibitem{bard_si}
        W.A.Bardeen,O.Piguet and K.Sibold, \pl{72}{231}{77};
\bibitem{landau}
       A.Blasi O. Piguet and S.P.Sorella, \np{356}{154}{91};
\bibitem{low}
        W.Zimmermann, \cmp{39}{81,90}{74};
        J.H.Lowenstein, \cmp{47}{53,68}{76}, \np{96}{189,208}{75};
        T.E.Clark and J.H.Lowenstein, \np{113}{109}{75};
\bibitem{kraus}
        E.Kraus and K.Sibold, Nucl.Phys.Proc.Suppl.51C(1996)81;
        E.Kraus and K.Sibold, Z.Phys.C68 (1995) 331-344; 
\bibitem{maiso}
       P.Breitenlohner and D.Maison, \cmp{52}{11,38}{77}, \cmp{52}{39,54}{77}, 
        \cmp{52}{55,80}{77}; 
\bibitem{aoki}
       R. Kawabe  K.i. Aoki, Z. Hioki, M. Konuma , T. Muta, 
       Prog.Theor.Phys.Suppl.73:1-225,1982; 
\bibitem{msgras}
       P.A.Grassi, {\it Renormalization of Standard Model  in the 
         ``t'Hooft-background'' gauge by means of automatic symbolic 
         computation}, Ph.D. thesis, 1997, Milan University; 
       P.A.Grassi, {\it Feynman rules and   
         functional identities for BFM-SM}, in preparation;
\bibitem{cla}
        T.E.Clark \np{111}{134}{76}.

       


        









       
 





   
        







 

\end{thebibliography}
\end{document}